\documentclass[%
 preprint, 
 amsmath,amssymb,
 aps, physrev,
]{revtex4-2}

\usepackage{caption}
\usepackage{graphicx}% Include figure files
\usepackage{dcolumn}% Align table columns on decimal point
\usepackage{bm}% bold math
\usepackage{xcolor}

\begin{document}

\preprint{APS/---}

\title{The influence of solute induced memory on interface migration}

\author{C. W. Sinclair}
\affiliation{Department of Materials Engineering, The University of British Columbia, Vancouver, Canada}%Lines break automatically or can be forced with \\

\author{J. Rottler}%
\affiliation{Department of Physics and Astronomy and Stewart Blusson Quantum Matter Institute, The University of British Columbia, Vancouver, Canada}

\email{Contact author: chad.sinclair@ubc.ca}

\date{\today}

\begin{abstract}
Interface migration governs microstructural evolution during phase transformations and grain growth thereby dictating those material properties that depend on microstructure.  Recent work continues to highlight the rich range of behaviors exhibited by migrating interfaces and the complex connection between these behaviors and the underlying atomistic processes that determine coarse-grained interfacial mobility. For interfaces moving at low homologous temperatures and small driving forces, we show that significant non-Markovian effects can arise that invalidate commonly used analysis methods. Specifically, we demonstrate that solute can act as a source of such non-Markovian motion. In turn, we introduce a time-local (TCL) propagator approach to account for memory-dominated short and intermediate time interface dynamics. This approach extrapolates to the long time diffusive limit, enabling robust mobility estimates from simulation windows far shorter than those required to observe linear scaling directly. Comparison with solute-free boundaries validates the method and quantifies solute drag in the Cahn-Hillert sense, providing a route to extract drag coefficients and effective mobilities across a range of solute concentrations.  Our results demonstrate that analysis including memory is essential for connecting atomistic simulations to continuum models and offer a practical framework for studying interface kinetics in systems with slow internal processes.
\end{abstract}


\maketitle

%\linenumbers

\section{Introduction}

Interfacial mobility, a widely used coarse-grained parameter describing the kinetics of interface migration, is a key factor controlling microstructural evolution and, consequently, the macroscopic behavior of materials\cite{Rollett2004RecrystallizationEdition,Gottstein2009GrainEdition,Sutton2007InterfacesMaterials}. Considerable effort has been devoted to bridging atomistic and microstructural length scales by extracting mobilities from molecular dynamics simulations (see e.g. \cite{Olmsted2009SurveyMobility,Mendelev2001Grain-boundaryModels,Han2018Grain-boundaryApproach}) and attempting to incorporate them into predictive models for microstructure evolution predictions (see e.g. \cite{Guerdane2018Crystal-meltModeling, Suhane2023AtomisticallyGrowth}). However, recent insights into the complexity of interface kinetics demonstrate that significant challenges remain in achieving a fully reliable connection across these scales \cite{Han2018Grain-boundaryApproach,Qiu2024GrainRatchets}.  One place this is seen is in attempts to connect atomistic simulation results to coarse-grained continuum models for solute-interface interactions, particularly when solute-drag effects are involved \cite{Wang2025Solute-dragInterfaces}. 

The mobility of a given interface reflects the outcome of a hierarchy of sub-processes involving both individual and collective atomic motions \cite{Zhang2009GrainLiquids,Chesser2022Point-defectDiffusion,Sharp2018MachineBoundaries,Alexander2013ExploringTechnique,Duncan2016CollectiveTransformations}. Atomistic simulations reveal a wide variety of such mechanisms, depending on interface character, chemistry, temperature, and driving force\cite{Olmsted2009SurveyMobility}. Grain boundaries, for example, exhibit kinetic features reminiscent of supercooled liquids and glasses, including highly degenerate energy landscapes with multiple competing barriers for atomic rearrangements\cite{Zhang2009GrainLiquids,Sharp2018MachineBoundaries}.

Mobility is typically extracted from molecular dynamics (MD) simulations using either equilibrium fluctuations, analyzed through the capillary fluctuation method \cite{Hoyt2014AtomisticMobility,Hoyt2010FluctuationsSimulations, Foiles2006ComputationFluctuations,Trautt2006InterfaceWalk} or in the linear response regime, where a driving force is imposed and the resulting a steady-state velocity is measured \cite{Janssens2006ComputingBoundaries,Mendelev2013ComparisonMigration}. The former assumes overdamped Langevin dynamics, such that the interface center of mass undergoes diffusion, but is generally limited to high temperatures where mobility is appreciable. The latter is particularly useful at low temperatures or for sluggish interfaces, though large driving forces can alter the mechanisms of migration and lead to non-linear response \cite{Deng2011AtomisticMotion}

Both of these methods assume that interfacial motion is Markovian and memory-free. In reality, the hierarchy of relaxation times among microscopic processes ensures memory effects \cite{Mishin2023StochasticSystems}. Even if individual processes are Markovian, collective mobility becomes so only once faster processes have relaxed (see e.g. \cite{Swope2004DescribingTheory}). A broad distribution of relaxation times delays this equilibration, and memory grows more pronounced at low temperatures as timescales diverge. In polymer physics, for example, memory effects in the Onsager coefficients arise due to sub-diffusive intra-chain relaxation\cite{Bement2025ModelsTechniques}. Interfaces interacting with solute atoms provide a similar example in the solid state, as solute introduces additional relaxation processes that further complicate migration.  If these processes are not well separated in time from the global motion of the interface then memory effects will persist over times comparable to those that are attainable in typical atomistic simulations.

Memory commonly manifests as transient sub-diffusion, seen as deviations from pure diffusion at short to intermediate times, with the diffusive regime recovered only asymptotically at long times. Such behavior can persist over the limited times accessible to molecular dynamics meaning that straight forward extraction of mobility from such simulations can be challenging \cite{Wang2025Solute-dragInterfaces}. Recent simulations of grain boundaries in dilute Fe–C binary alloys provide a clear example of this phenomenon, revealing sustained sub-diffusive dynamics throughout the simulation window \cite{Wang2025Solute-dragInterfaces}.

The presence of non-Markovian memory renders the conventional memoryless overdamped Langevin description inadequate. However, generalized approaches developed in biochemistry provide effective corrections for systems that show memory \cite{Sayer2023CompactDynamics,Dominic2023BuildingSimulations}. These methods enable both extraction of memory kernels and extrapolation of short, memory-influenced trajectories to the long-time diffusive limit relevant for coarse-grained microstructure models. Here, we adapt such techniques to study the coupling between solute, interface migration, and memory using a simplified kinetic Monte Carlo (kMC) model of interface migration in the presence of interacting solute. Despite its simplicity, the model captures the essential physics at minimal computational cost. We first demonstrate the inconsistency of these systems with the capillary fluctuation method, and then present an alternative mobility extraction based on mean boundary motion. Finally, we apply a time convolutionless (TCL) propagator approach to learn the interfacial dynamics under the influence of memory.

\section{Methods: Discrete Two-Dimensional Model for Interface Dynamics in the Presence of Solute}

To investigate interface dynamics with tunable memory effects induced by solute, we adapt the lattice model of ~\cite{Wicaksono2013AInteraction}, conceptually similar to that used in~\cite{Wang2025Solute-dragInterfaces} as well as other recent work ~\cite{Mishin2023StochasticSystems,Mishin2023StochasticSystemsb}. This discrete two-dimensional representation allows explicit control over boundary stiffness, solute binding, and diffusion kinetics while remaining computationally efficient. The interface is represented by a connected line of lattice sites; it is initialized as a straight segment along $x$ at $z=0$. Periodic boundary conditions are applied along $x$ with length $L_x=100\,a$ ($a$ being the grid spacing for species motion), while the system is open along $z$.

In the absence of solute, the interface energy is given by a discrete Gaussian solid-on-solid (DGSOS) Hamiltonian \cite{Lapujoulade1994TheSurfaces,Weeks1979DynamicsGrowth},

\begin{equation}
    \mathcal{H} = \frac{\Gamma}{2} \sum_{\langle i,j\rangle} \big(h_i - h_j\big)^2,
\end{equation}

\noindent where $h_i$ is the height of interface site $i$ and $\Gamma$ is the interfacial stiffness. A flat interface ($h_i \equiv \text{const.}$) minimizes $\mathcal H$. At finite temperature the boundary roughens via local moves. The rate for an elementary boundary move is

\begin{equation}
    \mathcal{R}_B = \nu \exp \! \big[-\frac{\big(Q_B+\Delta\mathcal H\big)}{k_BT}\big]
\end{equation}

\noindent with attempt frequency $\nu$, hop barrier $Q_B$, and $\Delta\mathcal H$ the energy change due to the move (computed from $\mathcal H$).

\begin{figure}[!htb]
    \centering
        \includegraphics[width=0.8\textwidth]{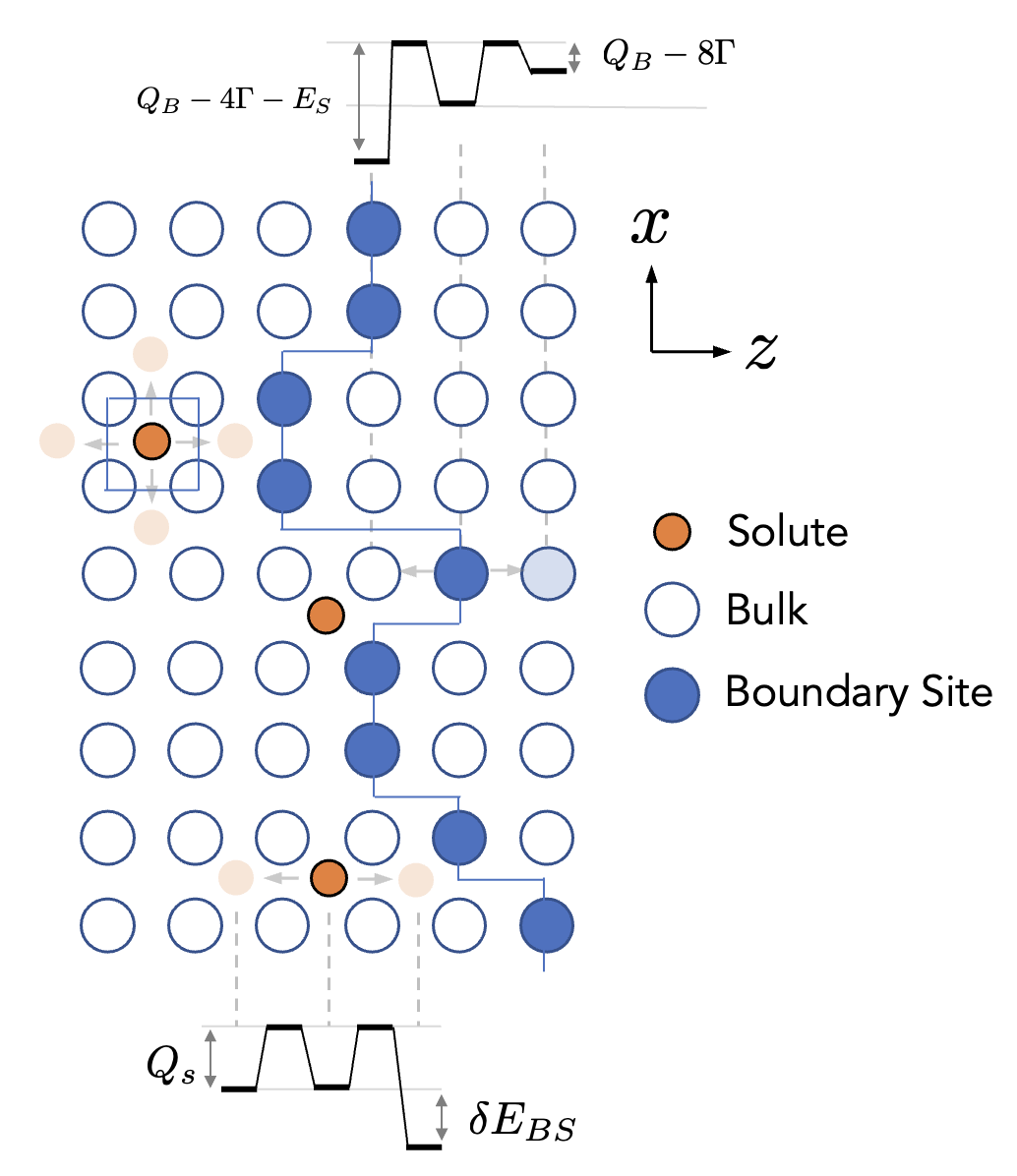}
    \caption{Schematic energy landscape for boundary and solute moves in the 2D DGSOS model. Blue sites denote interface sites, which can move to adjacent lattice sites (open circles). Solute atoms occupy interstitial positions between lattice sites and hop to neighboring interstitials sites if vacant. The top energy profile illustrates the landscape relevant to a boundary move. The bottom profile illustrates that for a solute move.}
    \label{fig:modelsetup}
    \end{figure}

Solute atoms occupy interstitial-like positions located between lattice sites (figure~\ref{fig:modelsetup}). They undergo excluded-volume diffusion to the four neighboring interstitials, with bulk hopping rate,

\begin{equation}
    \mathcal{R}_S = \nu \exp\!\big[-\frac{Q_S}{k_BT}\big].
\end{equation}

If a solute is adjacent to at least one interface site, two effects are included: (i) the barrier for the solute to hop away is increased by $\delta E_{BS}$, so the rate becomes $\nu \exp[-(Q_S+\delta E_{BS})/(k_BT)]$ and (ii) the system energy is lowered by a binding energy $E_S$ when solute and boundary are adjacent. These terms promote solute segregation to the interface and hinder boundary motion.
  
Dynamics are propagated using a residence-time kinetic Monte Carlo (n-fold way) scheme. At each step an event is selected with probability proportional to its rate, and time is advanced by $\Delta t = -\ln u / R_{\text{tot}}$ with $u\in(0,1]$ and $R_{\text{tot}}$ the sum over all rates. Trajectories for $h_i(t)$ and solute positions are resampled onto a uniform time grid for analysis (e.g., MSD and covariances).

We set $a=1$, $\nu=1$, and $Q_B=1$ to define the units of length, time, and energy, respectively. Thus quantities used here are reported as dimensionless ratios, e.g. the stiffness as $\Gamma a^2/Q_B$ and temperature as $k_B T/Q_B$. The parameters used in this work are summarized in Table~\ref{tab:parameters}.

\begin{table}[!htb]
    \centering
    \begin{tabular}{l c}
        \hline
        Parameter & Value \\
        \hline
        $Q_B$ (boundary hop barrier) & $1$ \\
        $\nu$ (attempt frequency) & $1$ \\
        $a$ (lattice spacing) & $1$ \\
        $\Gamma^\ast \equiv \Gamma a^2/Q_B$ (stiffness) & $1.5$ \\
        $Q_S$ (bulk solute hop barrier) & $Q_B$ \\
        $\delta E_{BS}$ (barrier boost near boundary) & $5\,Q_B$ \\
        $E_S$ (solute--boundary binding energy) & $5\,Q_B$ \\
        \hline
    \end{tabular}
    \caption{Model parameters for the DGSOS interface with solute.}
    \label{tab:parameters}
\end{table}

\section{Model Response}

We applied the model to two systems, one without solute and one containing 50 solute atoms. For the solute-bearing simulations, solute positions were initialized uniformly at random within $5a$ of the interface to accelerate segregation. At each temperature, we performed at least 10 independent runs of $5\times10^6$ kinetic Monte Carlo (kMC) steps. Each step executed a single event (either a boundary move or a solute hop), and the simulation clock was advanced according to the residence-time algorithm. The first $10^6$ steps were discarded for equilibration before kinetic measurements were collected.

\begin{figure}[!htb]
    \centering
    \includegraphics[width=0.5\textwidth]{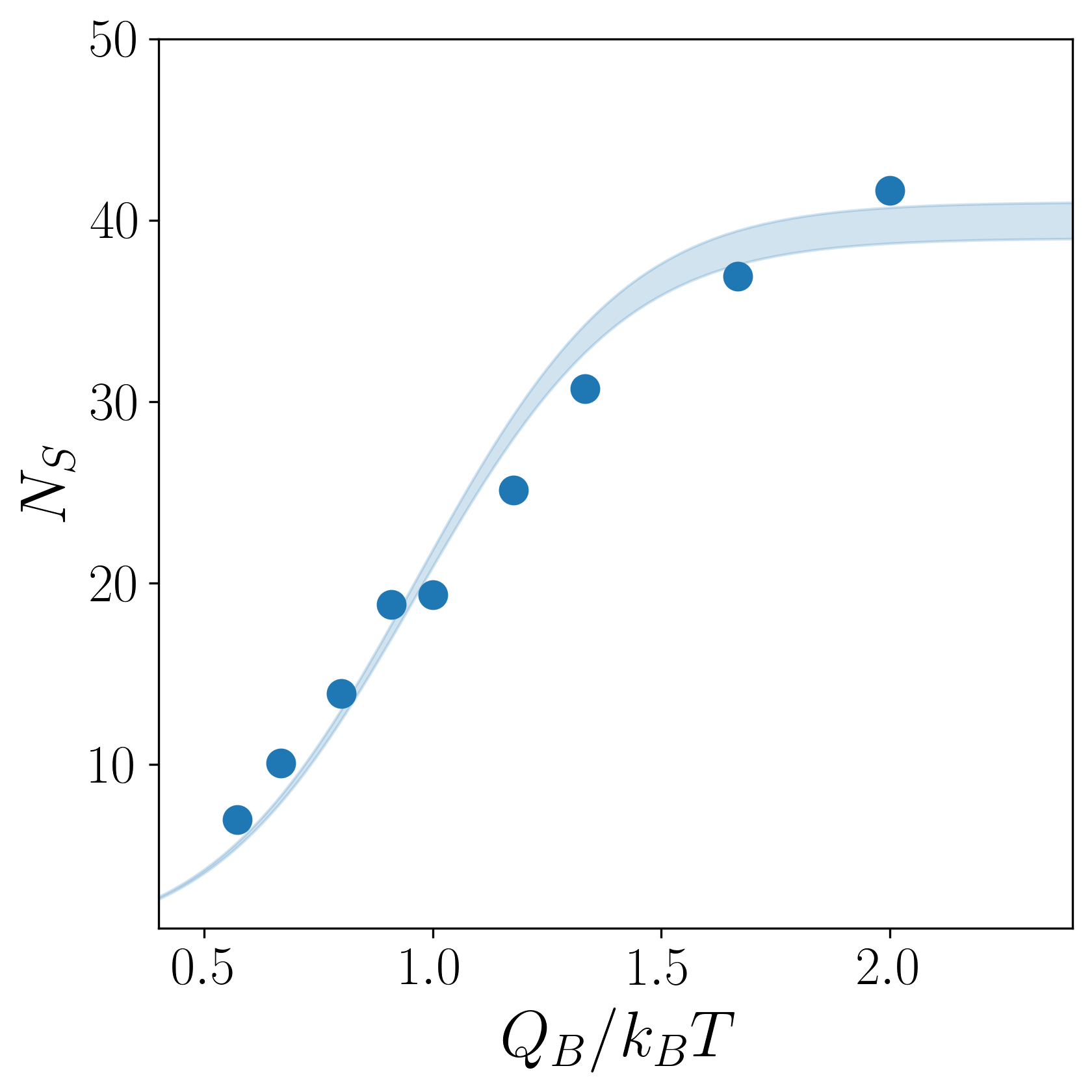}
    \caption{Equilibrated number of solute atoms occupying interfacial sites as a function of $1/(k_B T)$ (circles). The curve shows the finite-reservoir Langmuir–McLean prediction using the model energetics, with a correction to the total solute count to account for solute that diffuses away from the interface in the open $z$ direction.}
    \label{fig:NcvsKT}
\end{figure}

Figure~\ref{fig:NcvsKT} confirms that this equilibration period is sufficient for solute segregation to the interface. The equilibrated number of interfacial solute atoms agrees closely with the finite-reservoir Langmuir–McLean prediction. Because the system is open in $z$, a fraction of solute diffuses far from the interface. We therefore apply a correction to the total solute count in the isotherm to account for this loss to the bulk.

\begin{figure}[!hb]
    \centering
    \includegraphics[width=0.9\textwidth]{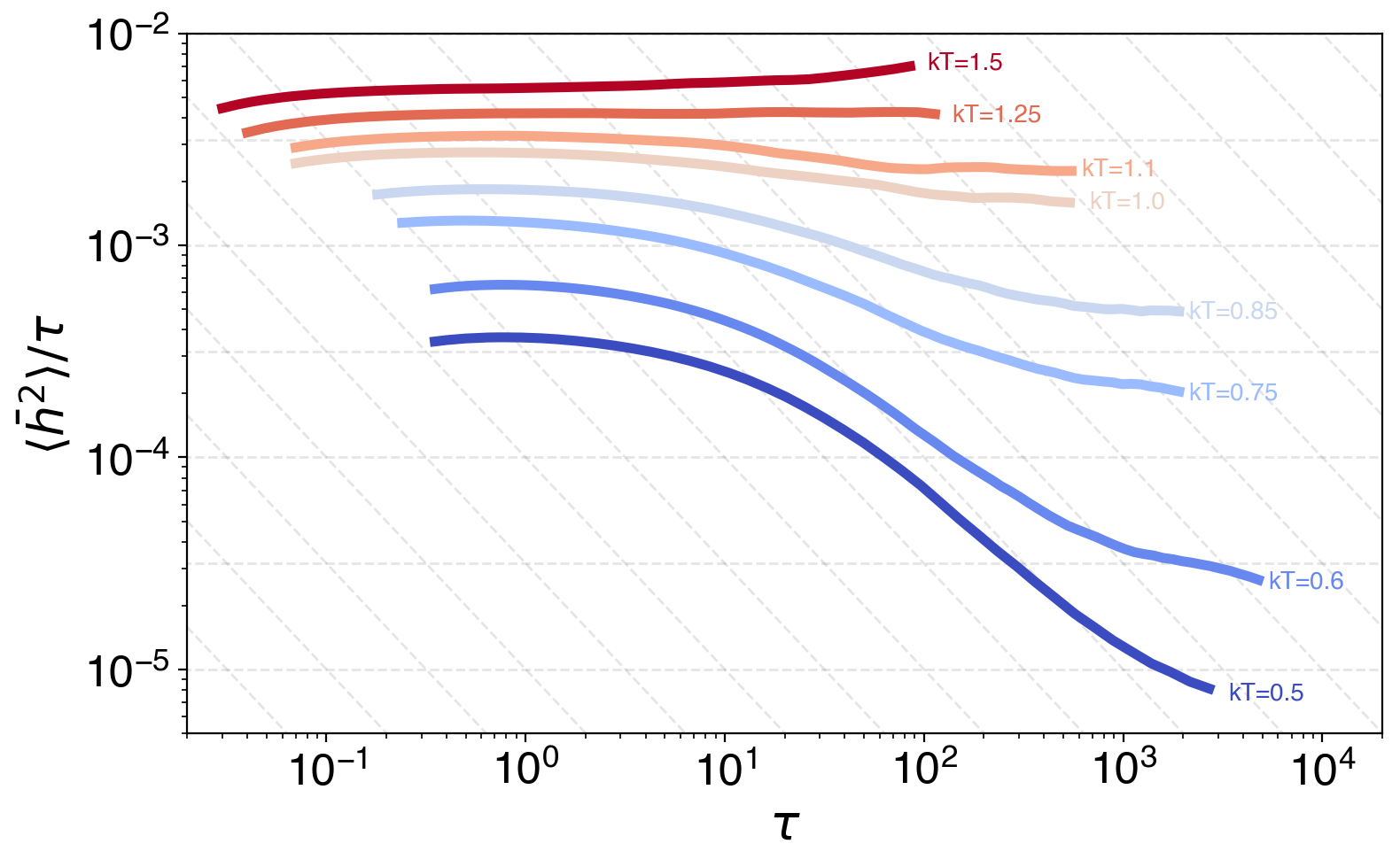}
    \caption{Mean-squared displacement of the mean interface position, normalized by lag time $\tau$, versus $\tau$ (log–log). A horizontal line corresponds to diffusive motion with magnitude $2D$. For $k_B T \gtrsim 1.1$, the response is diffusive over all lag times. At lower temperatures, subdiffusive transients appear at short lags and give way to diffusion only at long lags. At the lowest two temperatures the diffusive regime is not attained within the simulation time.}
    \label{fig:MSD}
\end{figure}

Figure~\ref{fig:MSD} shows the kinetics of interface motion for the solute-bearing systems as a function of temperature. We compute the mean-squared displacement as a time– and ensemble–average of increments of the mean interface position (resampled on a uniform time grid) over the post-equilibration window,
\begin{equation}
\mathrm{MSD}(\tau)\equiv \langle \Delta\bar{h}^2(\tau) \rangle \equiv
\big\langle [\bar h(t+\tau)-\bar h(t)]^{2}\big\rangle_{t} 
\qquad
\end{equation}

\noindent where $\langle\cdot\rangle_{t}$ denotes an average over all admissible time origins $t$ within each trajectory and over independent runs.
We plot the mean-squared displacement normalized by the lag time $\tau$ and displayed on log–log axes. A horizontal plateau indicates diffusive behavior with value $2D$, where $D$ is the diffusivity of $\bar h$. At high temperatures ($k_B T \gtrsim 1.1$), the curves are flat across all lags, indicating diffusive behaviour at all observed timescales. As temperature decreases, an initial sub-diffusive regime emerges (downward slope at short lags) before crossing over to diffusion at long lags. At the two lowest temperatures, the diffusive plateau is not attained within the simulation time window.

\section{Breakdown of the Capillary Fluctuation Method in the Presence of Solute}

At high temperatures, the kinetics of mean boundary motion exhibit simple diffusive behavior, consistent with the expected behaviour predicted by the Capillary Fluctuation Method. This can be demonstrated by starting from an overdamped Langevin description of the interface height profile, $h(x,t)$, as a function of position and time\cite{Hoyt2002AtomisticSystems,Hoyt2010FluctuationsSimulations,Hoyt2014AtomisticMobility,Foiles2006ComputationFluctuations}, 

\begin{equation}
\frac{1}{M}\frac{\partial h\left(x,t\right)}{\partial t} = \Gamma\frac{\partial^2h\left(x,t\right)}{\partial x^2} + \eta\left(x,t\right)
\label{eq:langevin}
\end{equation}

Here, $M$ is the mobility (inverse friction), the first term on the right-hand side represents a restoring force arising from the boundary stiffness $\Gamma$, and the final term $\eta$ corresponds to uncorrelated noise. Following the analysis of Trautt \emph{et al.} \cite{Trautt2006InterfaceWalk}, the mean boundary position can be written as:
\begin{equation}
    \bar{h}\left(t\right) = \frac{1}{L}\int_0^{L} h\left(x,t\right) dx
\end{equation}

Averaging the overdamped Langevin equation over space in a system periodic in $x$ eliminates the curvature term. The resulting dynamics reduce to a Wiener process, meaning that the mean interface position follows Brownian (diffusive) motion:
\begin{equation}
    \frac{d\bar{h}\left(t\right)}{dt} = M\bar{\eta}\left(t\right)
\end{equation}
Equivalently, the mean square displacement of an interface with area $A$ grows as, 
\begin{equation}
    \left<\bar{h}^2\right> = 2\frac{Mk_BT}{A}t = 2Dt
    \label{eq:msd}
\end{equation}

Equation \ref{eq:msd} defines the relationship between interface mobility ($M$) and diffusivity ($D$). For $k_BT > 1.1$, this relation is sufficient to determine $D$ (or $M$ via \ref{eq:msd}) by fitting a straight line to the mean square displacement versus time. At lower temperatures, it may still be possible to estimate the dynamics from the long-time behavior once diffusive limits are reached. However, in many cases this limit is never clearly observed, requiring an alternative approach.

A more general approach is to solve equation \ref{eq:langevin} directly. Following Hoyt\cite{Hoyt2002AtomisticSystems, Hoyt2010FluctuationsSimulations}, this can be done by applying a Fourier transform $h(x,t) \rightarrow \hat{h}_k(t)$, which converts the dynamics into a classic Ornstein–Uhlenbeck process for each $k$-mode:
\begin{equation}
\frac{d \hat{h}_k(t)}{dt} = - M \Gamma k^2 \hat{h}_k(t) + \hat{\eta}_k(t)
\end{equation}
From this, the equal-time covariance (or power spectrum) of the Fourier modes is obtained as,
\begin{equation}
\langle |\hat{h}_k|^2 \rangle = \frac{k_B T}{\Gamma}\frac{1}{k^2} 
\label{eq:variance1}
\end{equation}
providing a straightforward method to extract the interfacial stiffness. 

The time-lagged covariance, in contrast, is given by
\begin{equation}
\langle \hat{h}_k(t) \hat{h}_k^*(t + \tau) \rangle = \langle |\hat{h}_k|^2 \rangle \, e^{-M \Gamma k^2 \tau}
\label{eq:variance2}
\end{equation}
Importantly, these expressions are derived for the full covariance, implying that the results depend only on $k$. No mode coupling is expected, and the covariance matrices should therefore be purely diagonal, i.e. for $k \neq k'$ then $\langle \hat{h}_k(t) \hat{h}_{k'}^*(t + \tau) \rangle = 0$.

\begin{figure}[!htb]
    \centering
        \includegraphics[width=0.99\textwidth]{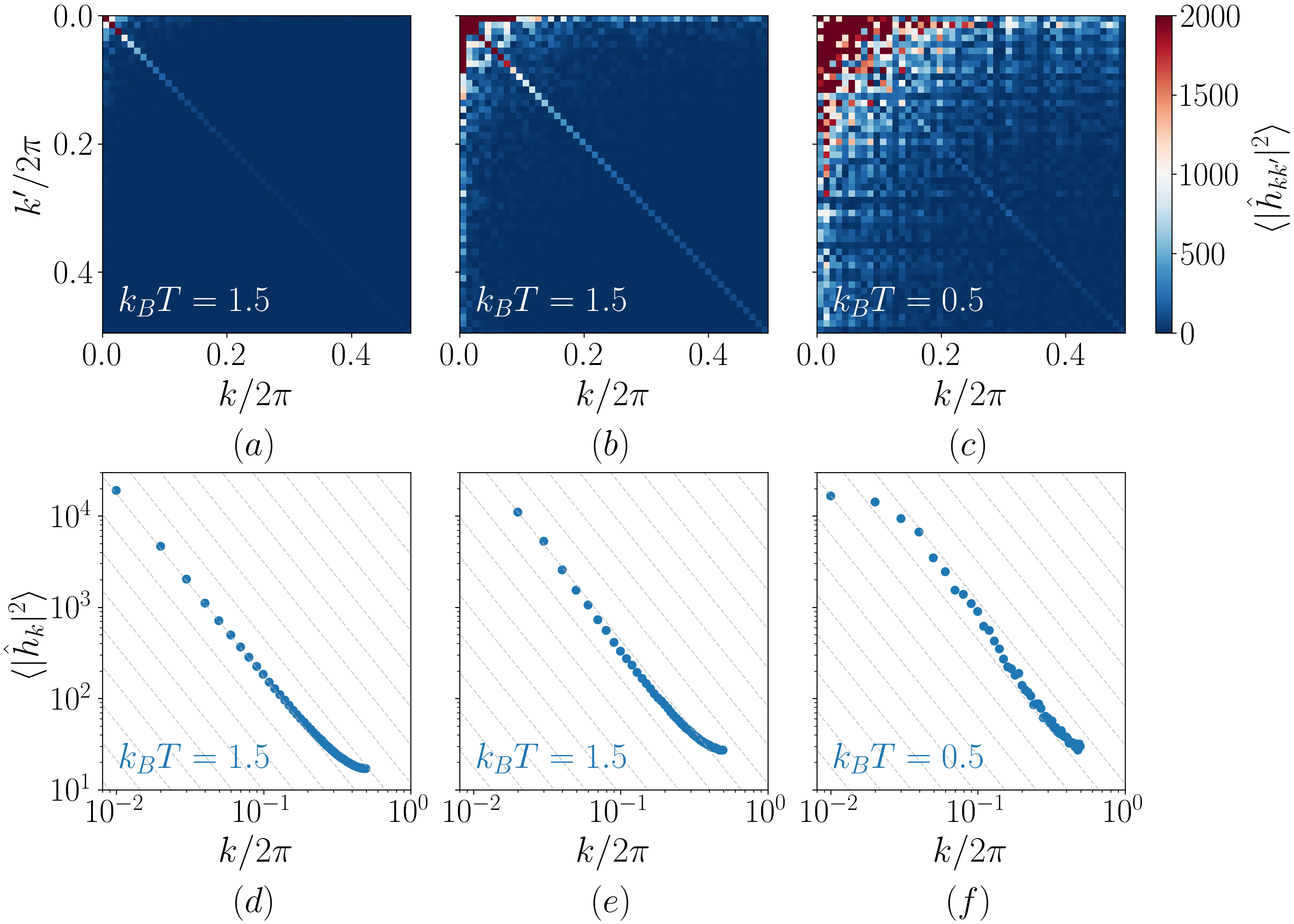}
        \caption{Autocovariance $\langle |\hat{h}_{kk'}|^2 \rangle$ for a) system with no solute at $k_BT$ = 1.5, b) system with 50 solute at $k_BT$ = 1.5 c) system with  50 solute at $k_BT$ = 0.5.  The solution to the overdamped Langevin equation \ref{eq:langevin} would predict only diagonal components along $k=k'$ while for the systems with solute off diagonal components (particularly at low $k$) start to appear, increasingly so as the temperature is lowered.  Figures (d)-(f) show the magnitude of the diagonal entries of the plots in the top row as a function of $k$ where equation \ref{eq:variance1} would predict a $k^{-2}$ dependence (slope = -2 on the graph as shown by the guidelines).  While this is well reproduced for systems without solute (Figures (a) and (c)), when solute is introduced deviations appear at low $k$ and at low temperatures the behaviour no longer is well predicted as being $k^{-2}$ dependent.}
        \label{fig:autocov}
    \end{figure}

Hoyt \cite{Hoyt2014AtomisticMobility,Hoyt2010FluctuationsSimulations} demonstrated that $\Gamma$ can first be determined from equation \ref{eq:variance1} and then used in equation \ref{eq:variance2} to extract both the interfacial stiffness and mobility. In our case, the stiffness is fixed by construction, so this method should, in principle, recover both parameters. Figure \ref{fig:autocov} presents the full auto-covariance matrix $\langle |\hat{h}_{kk'}|^2 \rangle$ under three conditions. In figure \ref{fig:autocov}a, for the solute-free system at the highest temperature ($k_BT = 1.5$), the data remain diagonal, and the diagonal components follow the expected $k^{-2}$ dependence of equation \ref{eq:variance1} (figure \ref{fig:autocov}d). When solute atoms are introduced at the same temperature, figure \ref{fig:autocov}b,e, deviations from strict $k^{-2}$ behavior appear at low $k$, corresponding to the emergence of off-diagonal components in the covariance matrix. Further lowering the temperature in the presence of solute enhances these off-diagonal elements and increases the deviation from $k^{-2}$ scaling, such that at $k_BT = 0.5$ the expected behavior is no longer observed (see figure \ref{fig:autocov}c,f). Unsurprisingly, similar mode-coupling effects also manifest in the time-lagged covariance matrix.

The appearance of off-diagonal elements in the auto-covariance matrix indicates that solute atoms cause coupling between Fourier modes, particularly at low $k$ (long-wavelength limit). This outcome is consistent with the underlying model. While uncorrelated fluctuations remain possible at large $k$, the presence of solute constrains long-wavelength modes, leading to their coupling.

When solute dynamics are fast compared to the interfacial modes of interest, the solute dynamics can be integrated out, yielding a renormalized free energy and a scalar stiffness. In this regime the equal-time spectrum is diagonal at sufficiently large $k$ and follows the behaviour expected based on equation \ref{eq:variance1}. In contrast, when solute dynamics are slow, the interface experiences a spatially non-uniform pinning field that couples Fourier modes. The covariance becomes non-diagonal and the $k^{-2}$ scaling breaks down at small 
$k$ so a single scalar stiffness ($\Gamma$) is not well defined.  In the solute-free model, translational invariance along the interface implies that correlations depend only on separations, yielding a diagonal covariance matrix. The introduction of solute creates a spatially varying pinning field that breaks translational invariance, so correlations depend on absolute position and different Fourier modes couple, the coupling being strongest at small $k$. Mode mixing is strongest at long wavelengths because the restoring term in the energy scales as $\Gamma k^{2}$ (vanishing as $k \to 0$) while the solute-induced couplings remain finite in the same limit. This means that the spatially heterogeneous pinning field can produce pronounced off-diagonal covariance at low $k$.

\begin{figure}[!htb]
    \centering
        \includegraphics[width=0.95\textwidth]{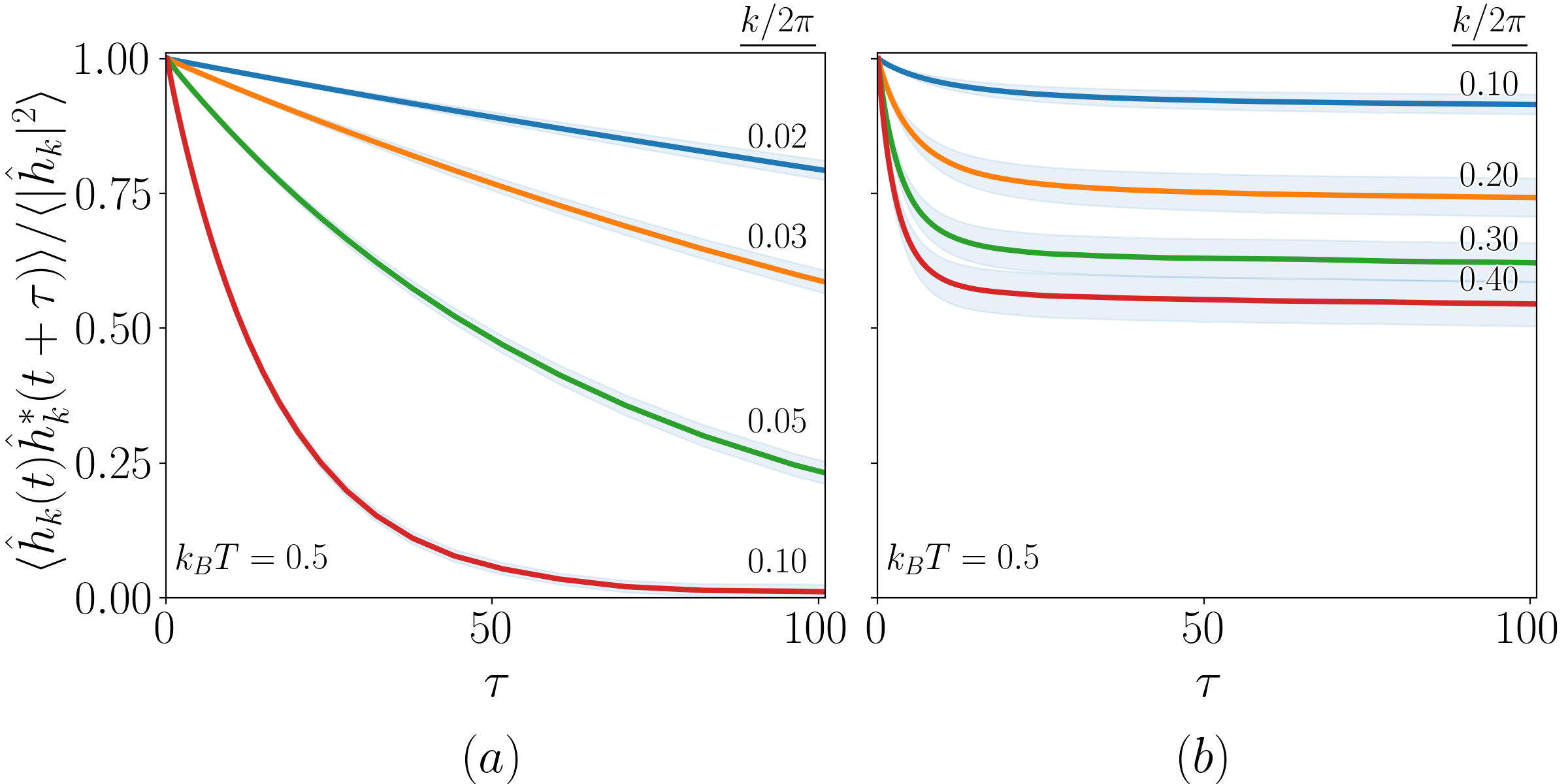}
        \caption{Normalized time-lagged covariance for $k = k'$ at selected values of $k$.  a) Results for system at $k_BT$= 0.5 without solute showing exponential decay as expected based on equation \ref{eq:variance2} b) system with solute at $k_BT$=0.5 showing non-exponential decay.  The shaded bands on both plots show the 99.5\% confidence interval, this illustrating the larger scatter in the system with solute, this corresponding to the larger scatter observed in figure \ref{fig:autocov}  Note that different $k$ values have been selected for a) and b) owing to the much faster decay in the system without solute.}
        \label{fig:laggedcov}
\end{figure}

As discussed above, mode coupling also affects the time-lagged covariance matrix and complicates the extraction of interface mobility. Figure \ref{fig:laggedcov}a shows the normalized time-lagged covariance for a solute-free system at $k_BT = 0.5$, which decays exponentially as predicted by equation \ref{eq:variance2}. In contrast, Figure \ref{fig:laggedcov}b presents the same analysis for an alloy system (50 solute) at $k_BT = 0.5$. Here, the decay deviates strongly from exponential behavior, and the scatter in the data is significantly larger than in the solute-free case, consistent with the increased scatter and mode coupling observed in figure \ref{fig:autocov}.

Because solute rearranges on a timescale comparable to or longer than that of interfacial relaxation at low temperatures, the interface dynamics inherit memory from this slow variable. This manifests as non-exponential decay in the time-lagged covariance (figure \ref{fig:laggedcov}b) and cannot be removed by basis changes such as would be done in a time-lagged independent component analysis (TICA) \cite{Schwantes2013ImprovementsNTL9,Perez-Hernandez2013IdentificationConstruction}. In the high-temperature limit, where solute motion is fast, the dynamics approach the Markovian, diagonal behavior predicted by equation \ref{eq:variance2}. In the low-temperature limit, where solute is slow, memory and mode coupling persist, rendering the standard Capillary Fluctuation Method inappropriate for extracting a single mobility $M$. Practically, this suggests either restricting analysis to regimes with fast solute dynamics or adopting a generalized (memory-kernel) description that accounts for mode coupling.  Generalized-Langevin (GLE) descriptions with memory kernels are widely used \cite{Tepper2024AccurateData,Dominic2023MemoryEfficiently}, but in practice they require an assumed functional form for the kernel and fitting to correlation data, an ill-posed, often non-unique inverse problem that must also satisfy fluctuation–dissipation.  Instead, we adopt another approach, specifically aimed at obtaining the mobility as the long-time limiting behaviour of the boundary motion.  We do this through the use of short simulations, on timescales where non-Markovian behaviour is still observed, but extrapolate this behaviour to the Markovian (diffusive) limit by learning the underlying non-Markovian dynamics \cite{Sayer2023CompactDynamics,Dominic2023BuildingSimulations}. This approach is described below.

\section{A time-local, memory enriched approach for extracting interface mobility}

The results above show that introducing interacting solute atoms creates two issues for the Capillary Fluctuation Method. First, mode coupling appears, making the extraction of a single scalar, coarse-grained stiffness ill-posed. Second, our focus here, the slow relaxation of solute kinetics, induces pronounced non-Markovian memory effects not present in the simple Langevin description. Rather than analyzing the full fluctuation spectrum along the interface, we concentrate here on the mean interface position, $\bar h(t)$, which should recover diffusive behavior in the long-time limit.  Our strategy is to bypass mode–coupling effects by focusing on the global coordinate $\bar h(t)$ and infer its long-time diffusivity from the decay of its scattering function, 
\begin{equation}
   \mathcal{C}_q(\tau)
   \;\equiv\;
   \big\langle \rho_q(t)\,\rho_q^{*}(t+\tau)\big\rangle_t
   \;=\;
   \Big\langle e^{\,iq[\bar h(t)-\bar h(t+\tau)]}\Big\rangle_t,
   \label{eq:CqDef}
\end{equation}

where $\rho_q(t) \equiv e^{i q\,\bar h(t)}$ is the Fourier–transformed density of the mean boundary position.
Here  $q$ is a small wavenumber chosen within the hydrodynamic limit (we reserve $k$ for in-plane Fourier modes used earlier). 
We then employ a time convolutionless (TCL), i.e. time–local, propagator to account for the memory–dominated regime at short and intermediate lag times, allowing us to extrapolate reliably to the Markovian diffusive limit.

In the Markovian (diffusive) limit, the increment $\Delta\bar h(\tau)=\bar h(t+\tau)-\bar h(t)$ has stationary Gaussian statistics with zero mean and variance $\langle[\Delta\bar h(\tau)]^2\rangle = 2D\tau$. Since $\mathcal C_q(\tau)$ is the characteristic function of $\Delta\bar h(\tau)$, we obtain
\begin{equation}
   \mathcal C_q(\tau)
   = \Big\langle e^{-iq\,\Delta\bar h(\tau)}\Big\rangle_t
   = e^{\Big(-\tfrac{q^2}{2}\,\langle[\Delta\bar h(\tau)]^2\rangle\Big)}
   = e^{(-D q^2 \tau)}.
   \label{eq:CqGaussian}
\end{equation}
Therefore, we can invert the above equation to relate $\mathcal{C}_{q}(\tau)$ to the mean square displacement as,
\begin{equation}
    \langle \Delta\bar{h}^2(\tau)\rangle = -2\frac{\log{\mathcal{C}_{q}(\tau)}}{q^2}
    \label{eqn:msdfromCq}
\end{equation}
and use this to define a lag–dependent diffusivity estimator
\begin{equation}
  \widetilde D(\tau)\;\equiv\;
  -\frac{\log \mathcal C_q(\tau)}{q^{2}\tau}
  \label{eq:Dtilde}
\end{equation}
so that $\widetilde D(\tau) \to  D$ as $\tau \to \infty$ in the hydrodynamic regime. 

Often, in other more complex systems comprised of many distinct and coupled processes an implied timescale (rather than a diffusivity) is computed.  In the present context this implied timescale would be defined as, $\tau_{imp}(q) = 1/(\widetilde{D}q^2)$ describing the relaxation time for each of the $q$ modes. If the process is memoryless and Markovian then $\widetilde{D}$ (and $\tau_{imp}$) are constant and independent of $\tau$.  Typically, for non-trivial models where there is no large time-scale gap between the slowest relaxation modes, it will take some time for memory to dissipate and for the system to become Markovian and for $\widetilde{D}$ (and $\tau_{imp}$) become constant.  

\begin{figure}[!htb]
    \centering
        \includegraphics[width=0.95\textwidth]{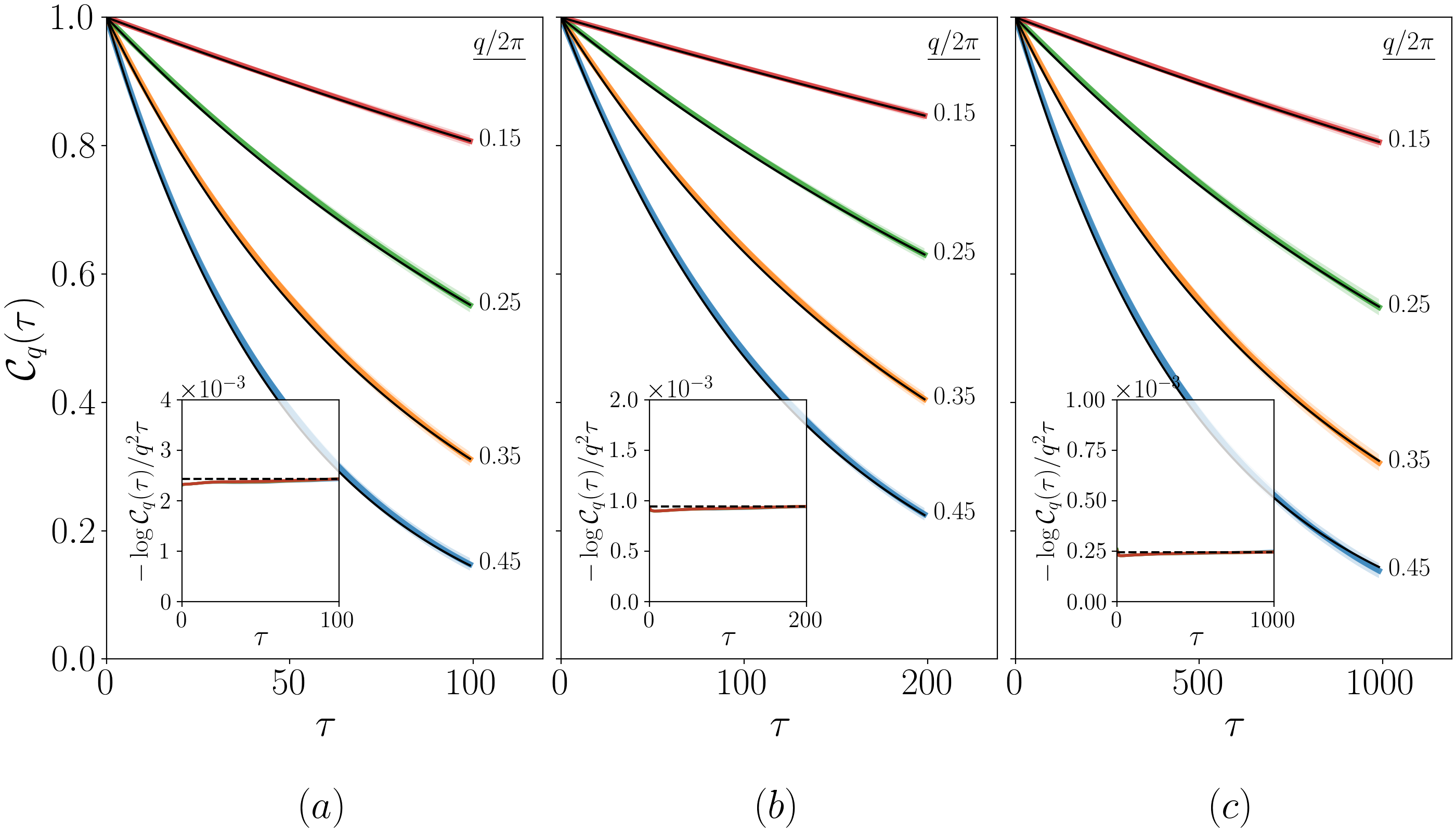}
        \caption{Decay of four selected modes, $q$ of the auto-covariance for interface without solute at a) $k_BT$=1.25 b) $k_BT$=0.75 and c) $k_BT$=0.5.  The inset to each figure shows that the results can be collapsed to give a nearly lag-time independent measure of the effective boundary diffusivity, $\widetilde{D}$, (dashed black line), this also showing that the decays are exponential from $\sim \tau$ = 0 suggesting memoryless Markovian behaviour.  The solid black line overlaid on each figure shows $e^{(-\widetilde{D}q^2\tau)}$ using $\widetilde{D}$ estimated from the insets.  Standard error of the mean has been calculated but is narrower than the thickness of the line in the main figures.}
        \label{fig:nosoluteAC}
\end{figure}

Figure \ref{fig:nosoluteAC} shows the evolution of $\mathcal{C}_{q}(\tau)$ for four different value of $q$ at three temperatures spanning $k_BT$=1.25, 0.75 and 0.5 for the \emph{pure system}.  In all cases the modes decay exponentially, this being seen most clearly from the plots of $-\log{\mathcal{C}_{q}(\tau)}/q^2\tau$ shown in the insets. These inset plots show that across the full temperature range studied the behaviour is well described by equation \ref{eq:CqGaussian} from $\tau \approx 0$. This compares well with expectation based on the observed linear variation of $\left<\bar{h}^2\right>$ with $\tau$ for systems without solute across all observation temperatures. The boundary diffusivities estimated from equations \ref{eq:Dtilde} and \ref{eq:msd} also match very well in this case.  

\begin{figure}[!htb]
    \centering
        \includegraphics[width=0.95\textwidth]{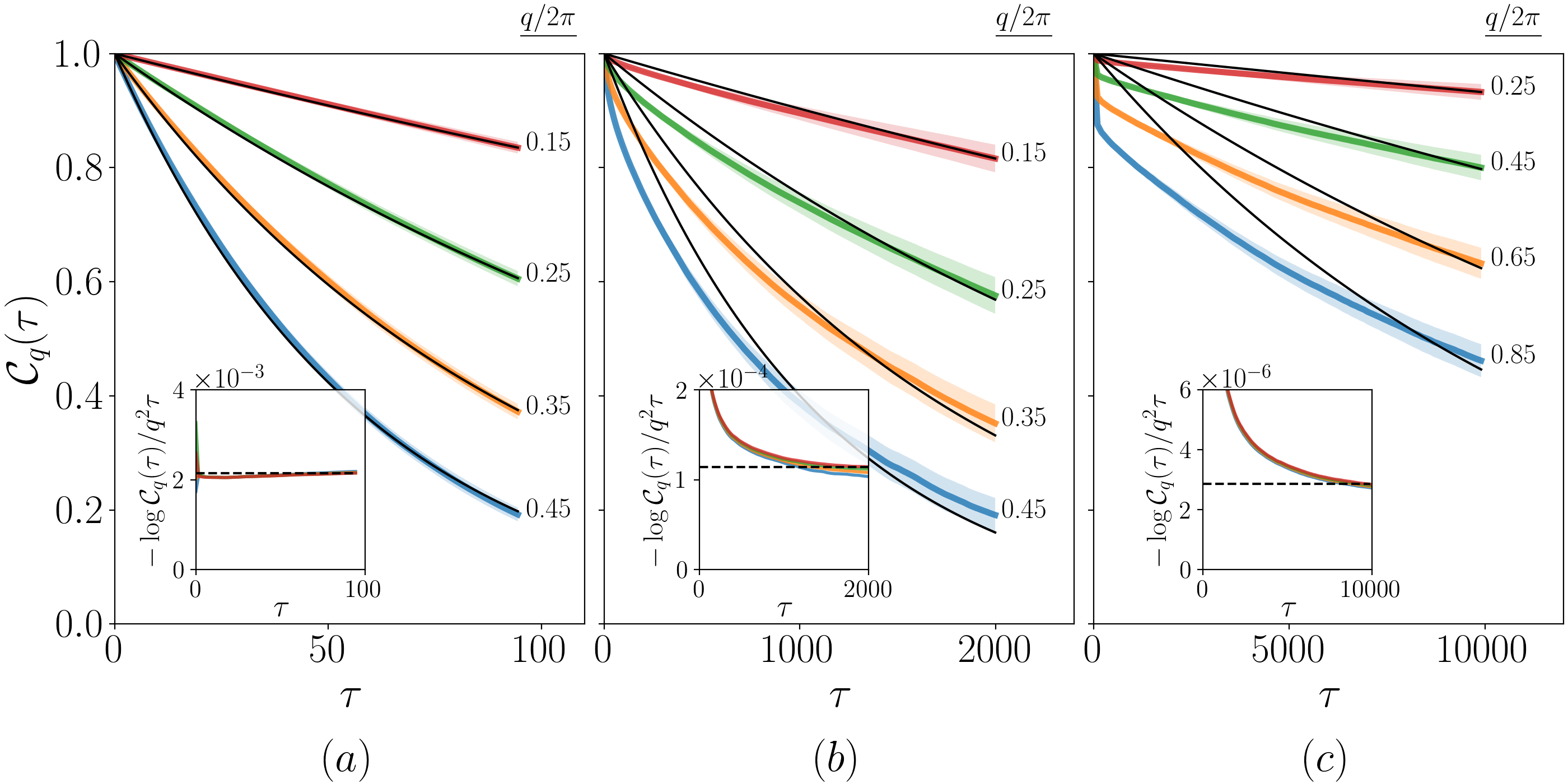}
        \caption{Decay of four autocorrelation modes, as in Figure \ref{fig:nosoluteAC},  but now for a system containing solute at a) $k_BT$ = 1.25, b) $k_BT$ =0.75 and c) $k_BT$ = 0.5.  Note that in (b) and (c) the evaluation times are $10\times$ longer than for the same temperatures compared to those in Figure \ref{fig:nosoluteAC} and that in (c) larger values of $q$ have had to be used to show similar decay to other plots.  As in Figure \ref{fig:nosoluteAC}, the solid black lines in the main figures corresponds to a plot of $e^{(-\widetilde{D}q^2\tau)}$ using the dashed black line in the inset as an estimate for $\widetilde{D}$. }
        \label{fig:soluteAC}
\end{figure}

Figure \ref{fig:soluteAC} shows the behaviour of the alloy system at the same three temperatures used in Figure \ref{fig:nosoluteAC}.  At the highest temperature ($k_BT$=1.25, Figure \ref{fig:soluteAC}a) the behaviour is very similar to that in Figure \ref{fig:nosoluteAC}a, the correlation decaying (nearly) exponentially from $\tau \approx 0$ and the inferred interface diffusivity being only slightly lower than that extracted from the system containing no solute at the same temperature.  This was observed for all temperature $k_BT > 1.1$.

At lower temperatures the effect of solute becomes apparent in multiple ways.  First, for the same set of $q$ values, much longer simulation times were required in figures \ref{fig:soluteAC}b and c to achieve similar levels of decay compared to the system without solute.  At $k_BT=0.5$ larger values of $q$ had to be used to visualize the decay compared to the system without solute as the slowing-down induced by solute was so strong. Also apparent in Figures \ref{fig:soluteAC}b and c is the non-exponential nature of the decays.  As shown in the insets, the behaviour across the full simulation time range does not converge to a constant value of $-\log{\mathcal{C}_{q}(\tau)}/q^2\tau$ indicating that we have yet to reach the diffusive limit.  

At low temperatures, slow solute relaxation induces memory effects that are usually represented by a memory kernel $\mathcal{K}(t)$ in a generalized approach \cite{Dominic2023BuildingSimulations}.  An equivalent but often more convenient representation is the time–convolutionless (TCL) form\cite{Sayer2023CompactDynamics,Dominic2023BuildingSimulations,Dominic2023MemoryEfficiently}, which rewrites the kinetics as a first–order equation with a time-local generator
\begin{equation}
  \frac{\partial}{\partial t} \big\langle\rho_q(t)\big\rangle
  = \mathcal{R}(t) \big\langle\rho_q(t)\big\rangle 
  \label{eq:TCLgen}
\end{equation}
where $\mathcal{R}(t)$ is the TCL generator whose explicit time dependence encodes the memory inherited from solute dynamics. Multiplying both sides by $\rho_q^{*}(t+\tau)$ and taking the ensemble average gives the corresponding relation
\begin{equation}
  \frac{\partial}{\partial t}
  \Big\langle \rho_q(t) \rho_q^{*}(t+\tau) \Big\rangle
  = \mathcal{R}(t)
    \Big\langle \rho_q(t)\,\rho_q^{*}(t+\tau) \Big\rangle
\end{equation}
For a stationary process we can shift the time origin and express this directly in terms of the lag variable $\tau$, leading to the compact form
\begin{equation}
  \dot{\mathcal{C}}_q(\tau) = \mathcal{R}(\tau)\,\mathcal{C}_q(\tau),
  \label{eq:TCLcorr}
\end{equation}
which makes clear that deviations of $\mathcal{C}_q(\tau)$ from pure exponential decay correspond to a lag–dependent decay rate $\mathcal{R}(\tau)$.  In the Markovian limit,
$\mathcal{R}(\tau)\rightarrow -Dq^2$ becomes constant, recovering
equation ~\ref{eq:CqGaussian}.

Direct evaluation of $\mathcal{R}(\tau)$ from equation~\ref{eq:TCLcorr} would, in principle, involve differentiating $\mathcal{C}_q(\tau)$, but this operation is sensitive to noise in simulation data\cite{Dominic2023BuildingSimulations}.  A more robust approach is to integrate equation~\ref{eq:TCLcorr}, which gives
\begin{equation}
    \mathcal{C}_q(\tau) = \mathcal{U}(\tau)\mathcal{C}_q(0)
    \label{eq:intTCL}
\end{equation}
where we define the propagator
\begin{equation}
    \mathcal{U}(\tau) = 
    \exp\!\left[\int_0^{\tau} \mathcal{R}(s)\,ds\right].
    \label{eq:Udef}
\end{equation}

Practically, $\mathcal{U}(\tau)$ can be obtained directly from the measured correlation function by inverting equation~\ref{eq:intTCL},
\begin{equation}
    \mathcal{U}(\tau) =     \mathcal{C}_q(\tau)\big[\mathcal{C}_q(0)\big]^{-1}
\end{equation}
Once the solute–induced memory has decayed beyond a finite lag $\tau_u$, the generator becomes constant, $\mathcal{R}(\tau\ge\tau_u)\rightarrow \mathcal{R}_\infty$, and the
propagator likewise approaches an exponential form,
\begin{equation}
    \mathcal{U}_\infty(\tau>\tau_u) = e^{\mathcal{R}_\infty \tau} 
\end{equation}
In this asymptotic regime the correlation function evolves as
\begin{equation}
    \mathcal{C}_q(\tau_u+n\Delta t)
    = \big[\mathcal{U}_\infty\big]^n\mathcal{C}_q(\tau_u)
    \label{eq:UpropRevised}
\end{equation}
with fixed lag increment $\Delta t$, providing a stable means of extrapolating to the Markovian diffusive limit without requiring long simulations.

Once the lag-time $\tau_u$ has been identified one can then evaluate both $\mathcal{U}_\infty$ and $\mathcal{C}_q(\tau_u)$ and thus predict the evolution of $\mathcal{C}_q(\tau)$ for all $\tau > \tau_u$.  Note that $\tau_u$ can be short relative to the times required for $\widetilde{D}$ to approach $D$, meaning that from relatively short trajectories, where the dynamics are still strongly influenced by memory, the limiting value of $\widetilde{D}$ can be obtained.  

\begin{figure}[!htb]
    \centering
        \includegraphics[width=0.95\textwidth]{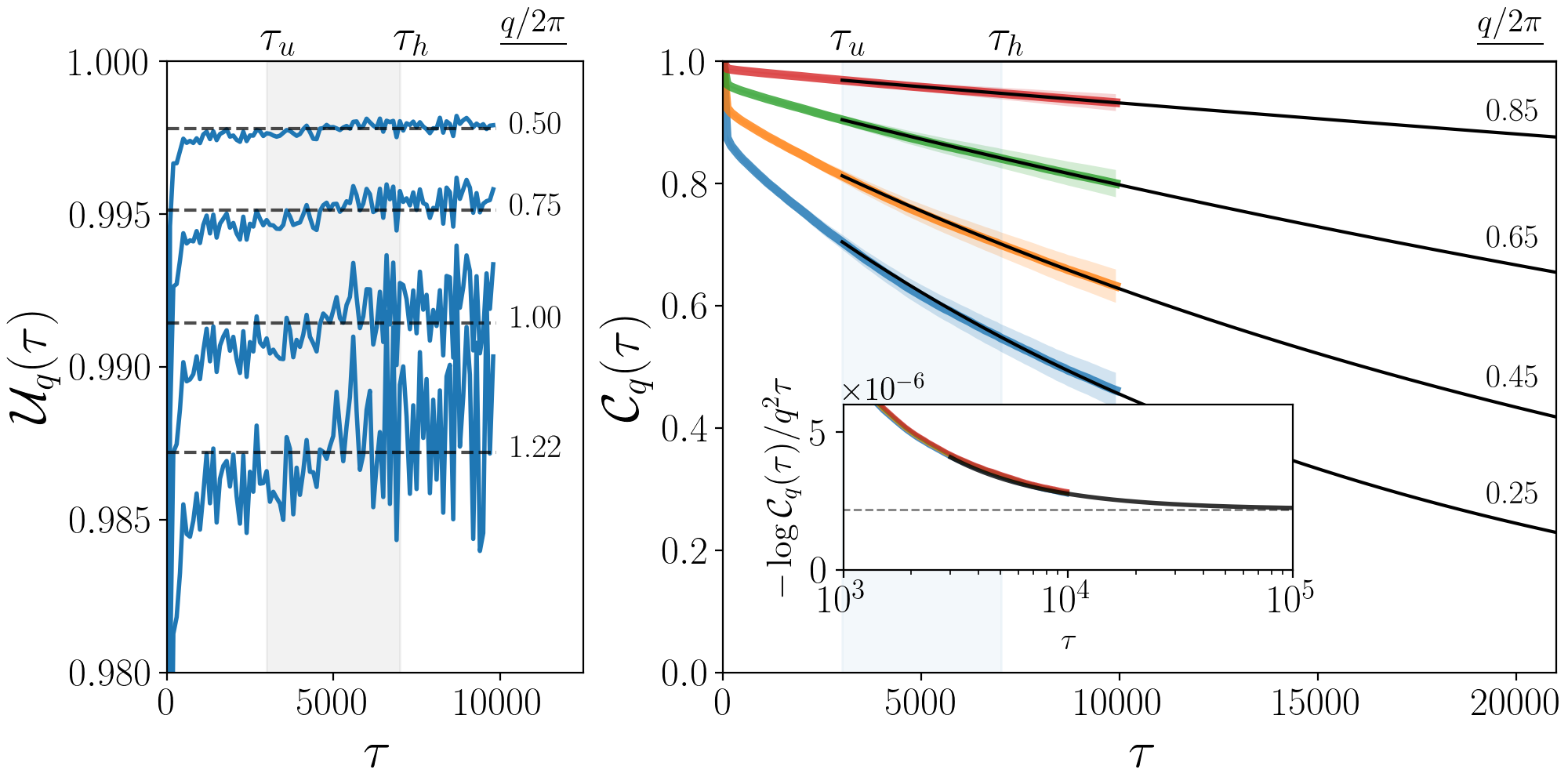}
        \caption{a) $\mathcal{U}(\tau)$ as a function of $\tau$ for four values of $q$ and data generated with solute at $k_BT$ = 0.5. As can be seen, $\mathcal{U}(\tau)$ tends towards a constant value at sufficiently high $\tau$.  To determine $\mathcal{U}_\infty$ we have selected a value of $\tau_u$ and calculated an average over a fixed window spanned by $\tau_u$ and $\tau_h$.  b) The $\mathcal{U}_\infty$ used in (a) was propagated via equation \ref{eq:UpropRevised} to generate the solid black lines, these being compared to the same data shown in Figure \ref{fig:soluteAC}.  The value of $\tau_u$ is shown from which $\mathcal{U}_\infty$ was propagated as well as the averaging window used (shaded region).  The inset shows $-\log{\mathcal{C}_{q}(\tau)}/q^2\tau$ which asymptotically approaches a long time estimate for $\widetilde{D}$.}
        \label{fig:TCL-kT0.5}
\end{figure}

Figure \ref{fig:TCL-kT0.5}a shows components at selected $q$ values for the propagator $\mathcal{U}(\tau)$ generated from the simulation performed at $k_BT$=0.5 with solute present (Figure \ref{fig:soluteAC}). As can be seen, the components of $\mathcal{U}(\tau)$ are noisy but do tend towards a constant value. Following \cite{Dominic2023BuildingSimulations}, we choose to compute $\mathcal{U}_\infty$ by averaging over a time window between the lower limit $\tau_u$, from which the dyanmics will be propagated using equation \ref{eq:UpropRevised}, and an upper limit $\tau_h$.  These limits are shown via the shaded box in this figure and the values of $\mathcal{U}_\infty$, taken as the averages within this window, are shown as the dashed black horizontal lines.  

Having identified $\tau_u$, the time at which we approximate true Markovian behaviour to commence, and $\mathcal{U}_\infty$ we can next use this information to propagate $\mathcal{C}_{q}(\tau)$ forwards from $\tau_u$ using equation \ref{eq:UpropRevised}.  The results of doing this are shown in Figure \ref{fig:TCL-kT0.5}b as the black solid lines in the main plot overlaid on the dynamics obtained from simmulation, the same data as shown in Figure \ref{fig:soluteAC}.  Again, the shaded box shows the windowed data used to generate $\mathcal{U}_\infty$, while the inset shows  $-\log{\mathcal{C}_{q}(\tau)}/q^2\tau$ computed both from simulation data as well as from the propagated dynamics from $\mathcal{U}_\infty$.  An estimate for $\widetilde{D}$ can then be obtained from extrapolation of the dynamics from propagating with $\mathcal{U}_\infty$, this being shown as the dashed line in the inset to Figure \ref{fig:TCL-kT0.5}b.

\begin{figure}[!htb]
    \centering
        \includegraphics[width=0.8\textwidth]{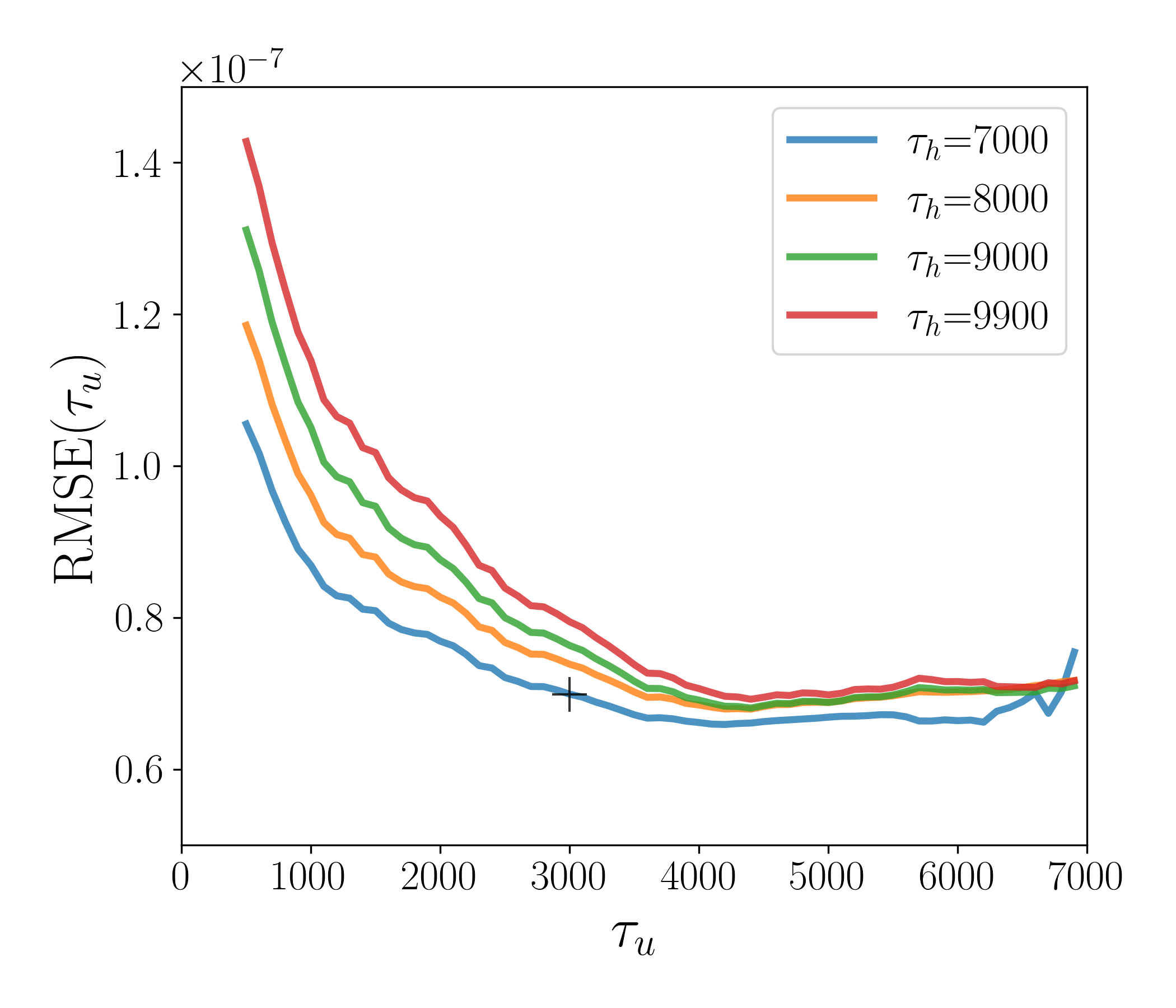}
        \caption{The root mean square error in $-\log{\mathcal{C}_{q}(\tau)}/q^2\tau$ calculated from the difference in $\mathcal{C}_{q}(\tau)$ taken directly from simulation versus that computed using the results from the TCL propagator $\mathcal{U}$ (equation \ref{eq:UpropRevised}) for the data shown in figure \ref{fig:TCL-kT0.5}.  Here, $\tau_h$ is taken as the upper limited used in the average to determine $\mathcal{U}_\infty$.  The condition marked by an `+' corresponds to the one used to generate the data in Figure \ref{fig:TCL-kT0.5}.}
        \label{fig:TCL-kT0.5-RMSE}
\end{figure}

The quality of the dynamic approximation using $\mathcal{U}_\infty$ is sensitive to the choices of $\tau_u$ and $\tau_h$.   Figure~\ref{fig:TCL-kT0.5-RMSE} quantifies this sensitivity by showing the root mean-squared error (RMSE) between the diffusivity-like quantities $\mathcal{D}_{C_q} = -\log{\mathcal{C}_q(\tau)}/(q^2\tau)$ (directly from simulation) and $\mathcal{D}_{U_q} = -\log{\mathcal{U}_q(\tau)}/(q^2\tau)$ (from the TCL-propagated dynamics), defined as

\begin{equation}
    \text{RMSE}(\tau_u) =  \frac{1}{N}\sum_{\tau>\tau_u}\left(\left[\mathcal{D}_{C_q}(\tau) -
      \mathcal{D}_{U_q}(\tau)\right]^2\right)^{1/2}
\end{equation}

Values of $\tau_u=3000$ and $\tau_h=7000$ were selected for the data in figure~\ref{fig:TCL-kT0.5-RMSE}, indicated by the `+' markers on the plot. This choice provides a balance; $\tau_u$ is large enough to suppress memory effects, while $\tau_h$ is small enough to maintain good statistics before noise dominates while also limiting the maximum computational effort necessary to estimate the diffusivity.

\begin{figure}[!htb]
    \centering
        \includegraphics[width=0.95\textwidth]{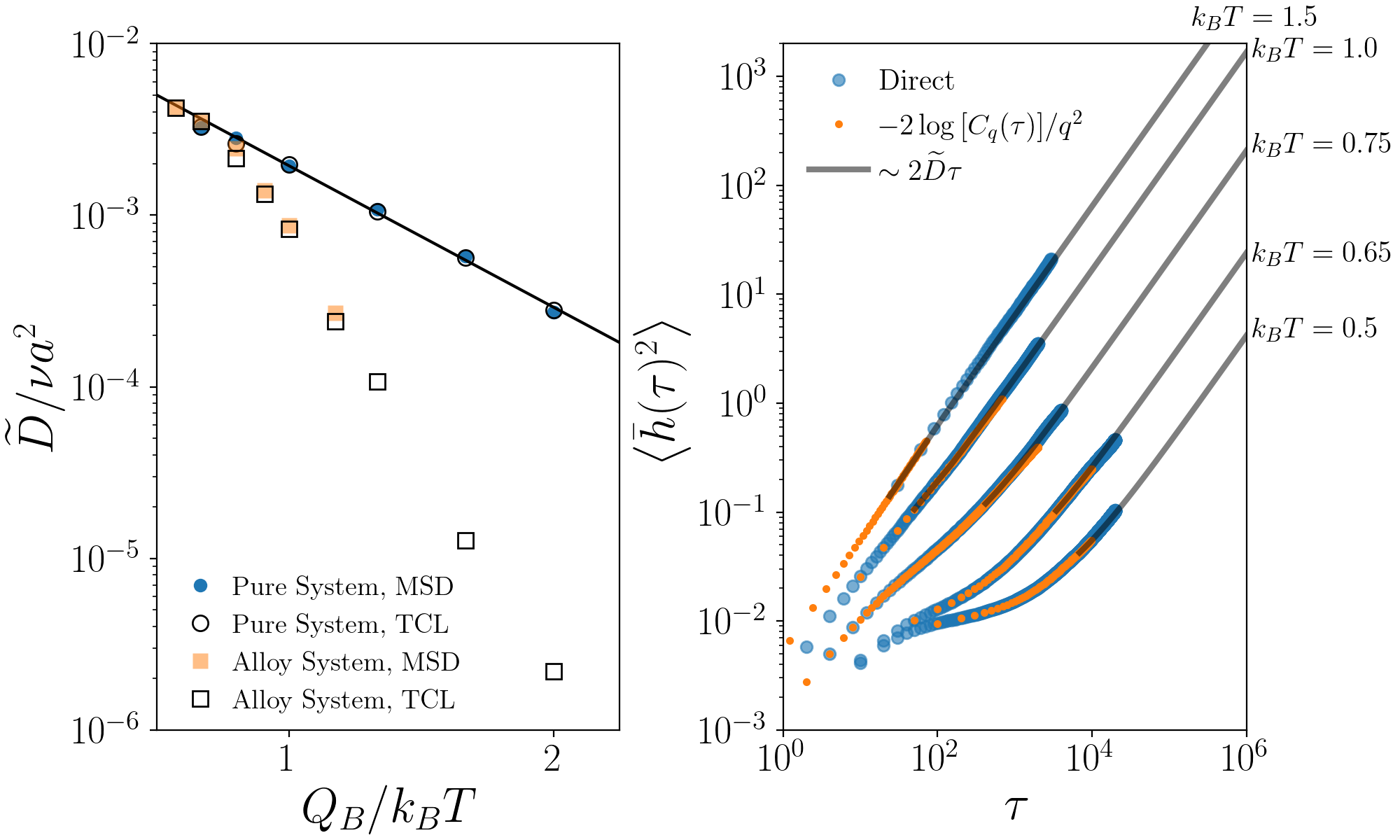}
        \caption{(a) Diffusivities determined for the pure and alloy systems. Closed symbols correspond to direct MSD measurements, while open symbols use the TCL-propagator approach (equation~\ref{eq:UpropRevised}). For the alloy at the lowest temperatures, no direct MSD-based measurement is possible because the diffusive limit is not reached. (b) Mean-squared displacement versus time for the alloy system at selected temperatures. Blue circles are direct MSD values, orange circles are reconstructed from $\mathcal{C}_q(\tau)$, and black lines show the diffusive limit inferred from $\widetilde{D}$ via TCL extrapolation.}
        \label{fig:diffusivities_fin}
\end{figure}

By selecting $\tau_u$ and $\tau_h$ in this way, we can extract the long-time diffusive dynamics, quantified through $\widetilde{D}$, from trajectories that are much shorter than those required to observe direct linear scaling of the mean-squared
displacement. We applied the same TCL-propagator procedure at all simulated temperatures to approximate $\widetilde{D}$ for both alloy and pure systems, as summarized in figure ~\ref{fig:diffusivities_fin}a. For comparison, diffusivities obtained directly from linear fits to the mean-squared displacement versus lag time are also shown where available. For the pure system, direct MSD fits are possible at all temperatures, as the dynamics are diffusive for all $\tau \ge 0$ and $\mathcal{C}_q(\tau)$ decays exponentially. For the alloy system, direct fits are impossible at the three lowest temperatures, and for $k_BT=0.85$, 1.0, and 1.1 linearity emerges only at long lag times. Nevertheless, the TCL-based diffusivities and the directly fitted values show excellent agreement across all conditions where both are available.

The TCL–propagator approach can also be used to \emph{predict} the long-time behavior of the system, not merely to infer $\widetilde{D}$. This is illustrated in figure ~\ref{fig:diffusivities_fin}b. The blue circles show the directly measured mean-squared displacement of the mean boundary position, while the orange circles represent values reconstructed from $\mathcal{C}_q(\tau)$ using equation~\ref{eqn:msdfromCq}. Starting from $\tau_u$, the black lines extrapolate the asymptotic diffusive regime using,

\begin{equation}
\langle \bar{h}^2(\tau > \tau_u) \rangle \approx  -\frac{2}{q^{2}}\log{\big(U_q(\tau_u)\big)}
+ 2\widetilde{D}\,\tau 
\end{equation}

demonstrating that the TCL method not only reproduces direct MSD data where available but also extends reliably into time windows where brute-force simulations cannot reach diffusive scaling within practical run lengths.

Beyond estimating $\widetilde{D}$, this framework could be extended to recover more detailed dynamical information. For instance, one could employ classical self-consistent memory–kernel expansions to invert $\mathcal{C}_q(\tau)$ and reconstruct the memory
kernel itself~\cite{Dominic2023BuildingSimulations}, thereby enabling calculation of the interface dynamics across \emph{all} timescales. Such extensions, while beyond the scope of the present work, illustrate the broader potential of combining TCL propagation with established non-Markovian analysis tools to bridge atomistic simulations and coarse-grained interface models.

\section{Discussion}

Our work builds directly on the study of Wang and Upmanyu~\cite{Wang2025Solute-dragInterfaces}, who demonstrated, using both simplified kMC  (similar to those presented here) and molecular dynamics simulations, that the presence of solute produces pronounced, transient sub-diffusive behavior at low temperatures and short simulation times. In their interpretation, this was modeled as an additional force term in the Langevin description, analyzed primarily in the short-time limit where solute motion was effectively frozen.  In contrast, we adopt a different perspective. Rather than treating the solute as a static force, we view its influence as a time-dependent friction (inverse mobility) encoded by the system’s memory. From this viewpoint, the classical concept of interfacial mobility, and the associated Cahn–Hillert \cite{Cahn1962TheMotion,Hillert1976AAlloys} solute-drag picture, is not rigorously defined until the slowest relaxation process (here, solute diffusion at low temperature) has equilibrated and the interface dynamics have become Markovian. Even after this point, residual memory effects continue to distort the dynamics, such that strictly linear behavior is recovered only asymptotically. Similar behavior has been reported in other contexts. For example, Rottler and M\"uller~\cite{Rottler2020KineticModeling} showed that even a short-ranged memory kernel can significantly influence microstructural evolution in polymer directed self assembly, long after apparent Markovianity is reached. Our results further suggest that the classical Langevin description may be inadequate for systems with slow solute kinetics; the breaking of translational invariance along the interface (evident from figure~\ref{fig:autocov}) corrupts the notion of a constant interfacial stiffness and complicates the application of a single, coarse-grained mobility parameter.

The non-Markovianity observed here is not restricted to interfaces interacting with solute atoms. In reality, interface motion is governed by a hierarchy of internal
relaxation processes that must equilibrate on timescales shorter than the motion of the mean interface position.  Ideally, these internal degrees of freedom (ranging from single, uncorrelated atomic hops to collective multi-atom rearrangements) would relax much faster than the boundary’s overall migration.  However, recent studies suggest that this separation of timescales does not always hold~\cite{Soltanibajestani2023AtomisticEnvironments}.   Molecular dynamics simulations of grain boundaries in pure BCC Fe, for example, reveal no strong timescale separation between internal and global dynamics, meaning that memory effects can persist for durations comparable to or longer than typical MD simulation windows. These effects become more pronounced as temperature decreases.  Furthermore, conditions below the roughening transition or near structural transformations within the boundary may themselves introduce slow collective processes. Such processes would extend the memory of the system far beyond the relaxation time of any single atomic mechanism, ensuring that non-Markovian effects persist to long times.  
Under these circumstances, mobilities extracted from short MD trajectories, assuming linear response and Markovian behavior, are unlikely to represent the true long-time dynamics reliably.

Finally, we return to the point emphasized by Wang and Upmanyu~\cite{Wang2025Solute-dragInterfaces} in the context of solute drag.  As in their study, we interpret the results in figure~\ref{fig:diffusivities_fin}a as a direct measure of interfacial
solute drag in the Cahn–Hillert \cite{Cahn1962TheMotion,Hillert1976AAlloys} sense. In the low (here effectively zero) net–velocity limit, the difference between the diffusivities, or the corresponding mobilities, of the pure and alloy systems measured at the same temperature represent the solute drag, consistent with Cahn’s original 1962 analysis \cite{Cahn1962TheMotion}. Cahn showed, using approximation and analytical expansion, that the
mobility of a solute–loaded boundary $M$ is related to the intrinsic mobility of a pure boundary $M_i$ via $M^{-1} = M_i^{-1} + \alpha C_b$ where $C_b$ is the bulk solute concentration and $\alpha$ is a parameter that depends on factors such as the solute binding energy profile and diffusivity. Thus, our calculations provide a means of extracting $\alpha$ in the low velocity regime by evaluating this mobility difference across a range of $C_b$ values.  

The method is not limited to zero–velocity interfaces. For driven boundary migration, the same approach can be applied provided dynamic equilibrium segregation is achieved and a steady drift velocity is maintained. By subtracting the net drift contribution from the mean boundary displacement, one can isolate the effects of velocity on solute drag. The main challenge in this case is ensuring that the solute distribution equilibrates dynamically within the short time windows accessible to, for example, molecular dynamics simulations. Developing or applying methods capable of bridging these timescales would therefore be highly valuable for extending the present approach to driven interface motion.

Together, these findings demonstrate that non-Markovian dynamics are not an isolated feature of solute drag but a general aspect of interface migration, underscoring the need for memory-aware mobility extraction methods in simulations of interfaces.

\section{Conclusions}

We have shown that the introduction of solute atoms into a simple two-dimensional interface model generates pronounced non-Markovian effects that contradict the assumptions underlying the capillary fluctuation method. By focusing on the mean interface position and employing a time–local (TCL) propagator formalism, we demonstrated how the long-time diffusive limit of boundary motion can be reliably extracted from relatively short trajectories that remain strongly influenced by
memory. This approach provides a consistent means of defining interfacial mobility in systems where slow relaxation processes, such as solute dynamics, dominate the kinetics.

Our analysis highlights two broader conclusions.  First, the classical Langevin description with constant stiffness and Markovian mobility is insufficient when translational invariance is broken and internal processes relax on comparable timescales to the mean boundary motion.  Second, the TCL–propagator method offers a practical strategy for bridging atomistic simulations with coarse-grained models, enabling
quantitative extraction of solute-drag coefficients and effective mobilities even in regimes where direct measurement is prohibitive.

\clearpage
\bibstyle{elsarticle-num-names}
\bibliography{GBMSM.bib}

\end{document}